# Connection between Bandgap Evolution and Strains of Octahedron in non-Perovskite $\beta$-MnO$_2$ under pressure: A First Principle Study

Li Li[1], Kuo Bao[1]*, Hui Xie[1], Youchun Wang[1], Xingbin Zhao[1], Xiaokang Feng[1], Hongyu Yu[1], Bingbing Liu[1] and Tian Cui[2,1]*

[1] *State key Laboratory of Superhard Materials, College of Physics, Jilin University, Changchun 130012, China*

[2] *Institute of High Pressure Physics, School of Physical Science and Technology, Ningbo University, Ningbo 315211, China*

# ABSTRACT

Lattice distortion due to octahedral rotation and distortion are high focussed, and it produces profound effect on a material's properties, such as its bandgap, magnetism and optical properties *etc*. Rutile-type *β*-$MnO_2$ is a wide used non-Perovskite magnetic material with octahedrons. We systematically studied its stability, electronic structures, magnetic structures, and optical properties within 0-100 GPa with density-functional theory (DFT). We find that the competition between bondlength and bonding angle leads its bandgap enlarging or shrinking within in the *Pnnm* phase with increasing pressure, because of the interaction of Mn-*d* and O-*p* states. We also find same pressure indused bandgap evolutions in the *Pnnm* phases of $SiO_2$, $GeO_2$, $SnO_2$ and $PbO_2$. The different ways of octahedral connection in *Pnnm* phase and *Pa$\overline{3}$* phase leads to an interesting pressure-induced bandgap enlarging. And in the *Pa$\overline{3}$* phase, the band gap can be tuned to 1.34 eV by pressure to meet the Shockley-Queisser limit. Moreover the two high pressure phases can be quenched to ambient pressure. Together with its mechanical, optical and antiferromagnetic properties, it ensures $MnO_2$'s application as a photovoltaic material for all working condition with multi-purpose. This study extends $MnO_2$'s application and give some new mechanism of pressure induced bandgap enlargement.

# I. INTRODUCTION

Octahedral ($AX_6$) rotation is important in solid state physics, which is significance to semiconductors and photovoltaic (PV) materials' application, particularly in the Perovskite materials [1]. It influences the bandgap width directly, thereby deciding their work efficiency. The competition between bond length and bond angel of octahedrons and the John-Teller effect modulate the bandgap in materials with octahedral structures competitively [2,3]. The connecting ways of octahedrons, such as edge-sharing, point-sharing and face-sharing within a material, also affect its bandgap. In the binary rutile oxides, such as $MnO_2$, $RuO_2$, $CrO_2$, $SiO_2$, $GeO_2$, $SnO_2$ $PbO_2$, *etc.*, they share same structures at a specific pressure range and contain $AO_6$ octahedrons. Moreover, due to their excellent chemical stabilities and appropriate bandgaps, they are widely studied [4,5]. Pressure, as an efficient and clean way, is extensive used, especially, to modify crystal structures and electronic structures. Therefore, to manipulate the octahedron by pressure is feasible. Meanwhile, the high-pressure phases process some intrinsic advantages, such as good mechanical properties, chemical inertia, thermal stability, radiation resistance, high thermal conductance, *etc* [6,7].

Manganese dioxide is highly focused for its wide applications in electromagnetic interfaces [8-12], high performance batteries [13], supercapacitors [14,15], catalyses [16,17], *etc.*, and it is abundant, cheap, and environmental friendly. As a semiconductor, if $MnO_2$'s bandgap is suitable, it could be a competitive PV candidate, that might overcome the disadvantages of the existing ones, such as expensive, unstable, inefficient, pollutive, *etc*. Theoretically studies shown that the Rutile-type $MnO_2$ with space group $P4_2/mnm$ ($\beta$-$MnO_2$) has a bandgap no matter based on the collinear or the noncollinear magnetic structures. Under negative pressure, it is half-metallic, however, the monolayer manganese dioxide is ferromagnetic (FM) semi-conductive with a high Curie temperature under pressure [18,19]. Some experimental work reported that $\beta$-$MnO_2$ is a screw-type structure with an extremely small band gap at low temperature with n-type carriers [20], which meets other measurements on its band gap about 0.27-0.28 eV [21]. There were experiments on the structural transitions of $\beta$-$MnO_2$ under high pressure (HP), which found that it transforms from a tetragonal phase ($P4_2/mnm$) phase to a orthorhombic (*Pnnm*) phase [22], then transforms to a cubic phase [23,24]. This also occurs in other Rutile-type dioxides under HP with interesting changes of their properties, such as driven $TiO_2$ the hardest material [25], and the blue shift of the optical properties of $SnO_2$ [26], and the second-order magnetic transition of $CrO_2$ [27], the ferroelastic transition of $GeO_2$ and $SiO_2$ [28,29]. However, as to the cubic phase of $\beta$-$MnO_2$, its space group is unknown, and the phase transition pressure is in disputation. Meanwhile, the HP properties of the $\beta$-$MnO_2$ is also little known.

With the above motivations, we present a systematic research on the structural, electronic, magnetic properties of $\beta$-$MnO_2$ under HP with DFT based on first-principle methods. We make a careful study on evolution of the bandgaps due to the distortion and rotation of octahedral with the pressure reported.

## II. COMPUTATIONAL METHODS

We searched for thermodynamically stable MnO$_2$ structures using the evolutionary algorithm with two and four formula units (f.u.) at the moderate pressure 0-100 GPa at zero temperature with the USPEX [30]. Our calculations were carried with spin-polarized DFT as implemented in the VASP [31]. The projector augmented wave method was adopted with $2s^22p^4$ and $3d^64s^1$ as the valence electrons for O and Mn atoms, respectively. We optimized the different geometrical structures (i.e., $P4_2/mnm$, $Pnnm$, $Pa\bar{3}$ and $R\bar{3}m$) and magnetic phases (i.e., NM, FM, AFM). After performing accurate convergence tests, a cut-off energy of 850 eV and the Mokhorst-Pack $k$-points meshes with a reciprocal space resolution of $2\pi \times 0.03$ Å$^{-1}$ were used [32], ensuring the total energy could be well converged to better than 1 meV/formula units (f.u.). The elastic constants were calculated with the strain–stress method. The bulk modulus B, and shear modulus G were estimated using the Voigt–Reuss–Hill (VRH) approximation [33]. In addition, to clarify the structural stability, we calculated the phonon dispersion curves with PHONOPY[29] code based on a supercell approach with the force-constant matrices.

For the description of the exchange and correlation (XC), we used the generalized gradient approximation (GGA-PBE). In order to overcome the deficiency of underestimate the band gap, we used the GGA+U method that introduces a $d$-$d$ intra-atomic Coulomb Energy U which is set to 3.1 eV base on the scheme of Dudarev et al [22]. In the scheme U is defined as $U_{eff} = U - J$, J is set to 0. And the full anisotropy approach, GGA+U+J, based on the scheme of Liechtenstein *et al* [34] was also applied. According to previous studies the values of U and J are set 4 eV and 0.9 eV, respectively [18,35]. Due to the results of GGA+U are similar with GGA+U+J, we only show the results of GGA+U in this paper.

## III. RESULTS AND DISCUSSION

We studied the thermodynamic and dynamical stabilities of the seven candidate HP-phases of MnO$_2$ with different magnetic states. Figure 1(a) shows the total energy as a function of the volume per unit cell with the GGA+U+J, the result of GGA is similar as the GGA+U+J [Fig. S1 in the Supplemental Material]. The spin-polarized calculations for the $R\bar{3}m$ structure always converge to the FM sates but the phonon spectrum is soft mode [Fig.S2 in the Supplemental Material]. The relative enthalpy-pressure curves in Fig. 1(b) shown that our predicted $P4_2/mnm$ phase is in well coincidence with the experimental result [36], which testified our calculations are credible. For the $Pnnm$ and $Pa\bar{3}$ phases, all the AFM-states are more stable than the FM-states. The AFM-$P4_2/mnm$ as ground state is stable with increasing pressure until 9.9 GPa. Above 9.9 GPa, MnO$_2$ undergoes a structural phase transition from the AFM-$P4_2/mnm$ phase to the AFM-$Pnnm$ phase, and then to a cubic-type structure AFM-$Pa\bar{3}$ phase above 44.7 GPa. These results clarify that the ambiguous high-pressure cubic state[23]. For the other four states (FM-*Pnnm*, AFM-*Pnnm*, FM-*Pa$\bar{3}$*, AFM-*Pa$\bar{3}$*), they are dynamic stability at 0 GPa [Fig. S3]. Therefore, the FM phases are metastable and could be quenched to ambient pressure. It might be FM and AFM for *P4$_2$/mnm*, *Pnnm* and *Pa$\bar{3}$* structures, six configures altogether. The optimized equilibrium lattice parameters of these phases are summarized in the Supplemental Material Table 1. The simulation of the 2.48 μB unpaired spin moment with GGA is closer to the experimental measurement of 2.35 μB [37] than the simulation value of 2.76 μB with GGA+U+J. For the two HP phases, lattice constants are consistent with the experiments [23]. While, we do not find any experiment on the magnetic moments to compare.

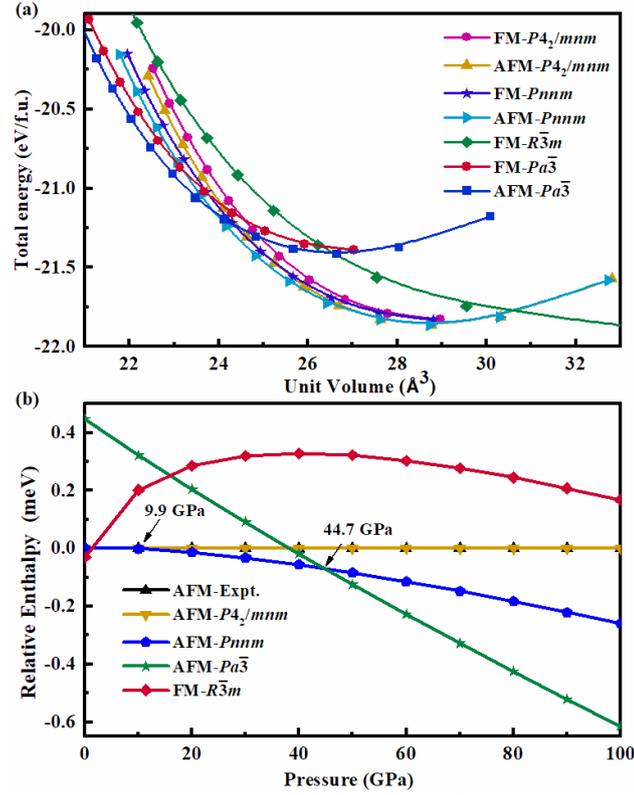

**FIG. 1.** (a) The calculated total energies per formula unit as a function of volume for the MnO$_2$ structures with GGA+U+J. (b) Computed relative enthalpy diagram of all considered MnO$_2$ structures relative to experiments on $\beta$-MnO$_2$ as a function of pressure.

The crystal structures of *P4$_2$/mnm*, *Pnnm*, and *Pa$\bar{3}$* phases of MnO$_2$ are presented in Fig. 2. As it could be seen that there is a distorted octahedron made up with six oxygen atoms. The four-oxygen-atom rectangle plane lies on the middle of the octahedron for all these three phases. The *P4$_2$/mnm* and *Pnnm* states are composed of a chain with the edge-sharing MnO$_6$ octahedrons, and the point-sharing octahedrons make up the *Pa$\bar{3}$* state. After a structural optimization on *P4$_2$/mnm* phase of AFM states with GGA and GGA+U+J, the lattice constants are 4.44, 2.87, 4.44 and 2.91 Å, respectively. These results agree with former experiments[20] and calculations [18,38]. We calculated the *Pnnm* and *Pa$\bar{3}$* AFM-phases' spin magnetic moments of the Mn-*d* and O-*p* as a function of pressure with both GGA+U+J (red fill circle and squares) and GGA (blue open circle and squares) [Fig. S4]. We find that the moment of *d*-orbital of Mn site decreases, and *p*-orbital of O increases with increasing pressure.

To understand physics of this interesting bandgap evolution, we systematically studied the partial density of states (DOS) for both spins computed of $\beta$-MnO$_2$ under HP to explore the origin of the physical properties for bandgap shifts. In order to obtain more accurate results, the GGA, GGA+U+J and WIEN2K were applied. We calculate the DOS using the full-potential linearized augmented plane-wave (LAPW) with RK$_{max}$ =

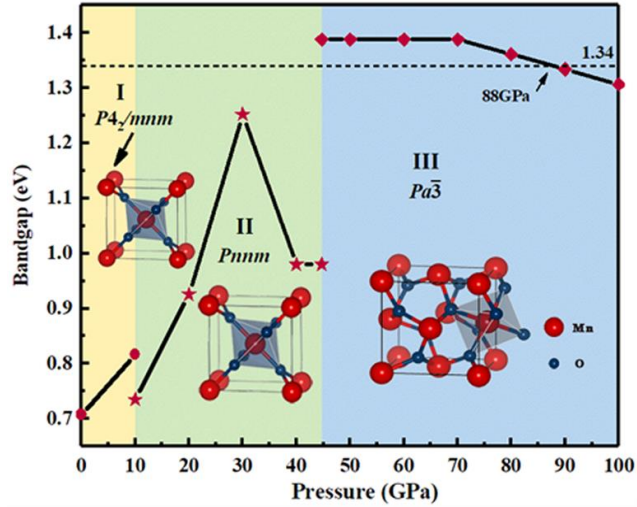

FIG. 2. The pressure dependence of the calculated band gaps of $MnO_2$ with antiferromagnetic state based on WIEN2K. The yellow, green, blue areas represent three different phases. The inset shows the crystal structures of $P4_2/mnm$, *Pnnm*, and $Pa\bar{3}$ phase of $MnO_2$. Large (red) spheres are manganese and small (blue) oxygen.

7.0 and muffin tin radii of 2.01 $a_0$ for Mn and 1.51 $a_0$ for O bases on WIEN2K [39]. The fully localized limit (FLL) double-counting correction [40] is applied for the localized *d*-orbitals, and the values of U and J are similar with GGA+U+J. For $P4_2/mnm$-phase at ambient pressure, neither the GGA results or the GGA+U+J results are metallic [Fig. S5 in the Supplemental Material]. The simulated DOS with GGA shows that Mn-*d* orbits dominate near the Fermi level ($E_F$) completely. However, this is incompatible with other theoretical results (the bandgap of PBE+U: 0.2 eV, WIEN2K: 0.8 eV, SCAN: 0.43 eV) [41] and experimental results [20]. On the contrary, when the *d-d* intra-atomic Coulomb energy U is taken into account in WIEN2K [Fig. 3], the Mn-*d* and O-*p* orbitals hybridize. Due to the mutual repulsion, Mn-*d* states are repelled above the Fermi level and O-*p* are pushed below the Fermi level, which manifests n-type semiconductor and coincides with the experiment [20]. We can clearly see in Fig.2 that the bandgaps expand for the three AFM-phases, for the high pressure phases, we can see different evolution patterns within them.

This gap is 0.7 eV at 0 GPa for the $P4_2/mnm$-phase which is consistent with experimental measurements [20]. In the *Pnnm*-phase, the band gap ranges 0.73 to 1.25 eV, and it increases firstly and then decreases with pressure increasing. The bandgap enlargement can be interpreted with the coupling O-*p* and Mn-*d* orbital, which comes from the decreased Mn-O-Mn bond angle [42,43]. Meanwhile, the octahedrons gradually rotate with pressure altering the γ angle, which is the reason for the bandgap expansion. Then, the shrinking of the bonds play a dominant role in the reduction of the bandgap, and the enhancement orbital overlap of Mn-*d* and O-*p* states is the reason. After transformed to the $Pa\bar{3}$-phase, the structure changes from edge-sharing and point-sharing octahedrons to point-sharing ones. It is the main reason for the abnormal bandgap enlargement between the *Pnnm*-phase and the $Pa\bar{3}$-phase. The gap is 1.39 eV at 70 GPa. At 88 GPa, it shrinks to 1.34 eV, which meets the Shockley-Queisser limit, the most desirable bandgap for PV materials. Moreover, both the GGA-PBE and the FLL results show that the gap enlargement is consistent. From Fig. 4, we can see that the Mn-O bonds of $P4_2/mnm$ phase along z axis perpendicular to the rectangle plane but the bond lengths are unequal. With pressure increasing, O atoms within $β$-$MnO_2$ shift slightly, the rectangle planes of the octahedron are not perpendicular to z axis. The distorted octahedrons rotate around the z axis gradually, and rotating directions are shown in Fig. 4. Consequently, the tetragonal $P4_2/mnm$ phase transforms to the orthorhombic *Pnnm* phase due to symmetry-breaking. Then, the orthorhombic phase undergoes a refractory phase transformation to the cubic $Pa\bar{3}$ phase, all Mn-O bonds within the octahedrons are equal in length and only the Mn-O-Mn angle have a tiny rotation around 90°. In the FM-$Pa\bar{3}$-phase, the dz$^2$, dx$^2$-y$^2$ + dxy, dxz + dyz are degenerate. While in the AFM phase, the degeneracy is removed and split into 5 nondegenerate energy levels, which could be explained with the Jahn-Teller effect. We conclude that AFM phase is more stable than the FM phase [Fig. S6 in the Supplemental Material].

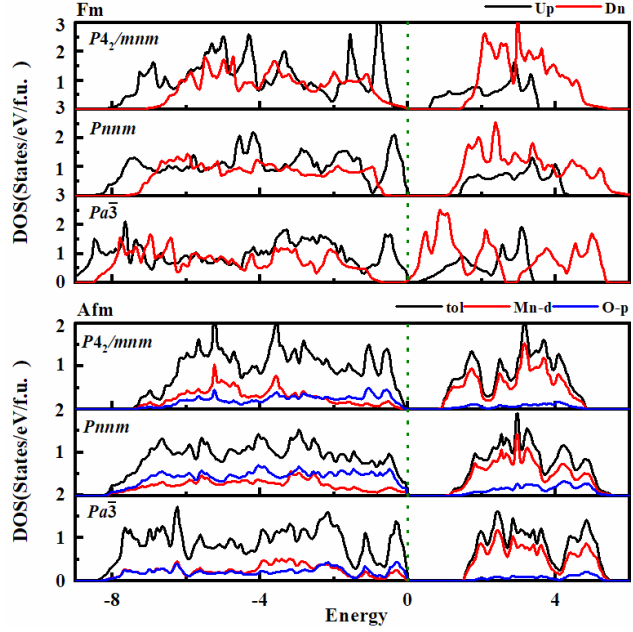

**FIG. 3.** Calculated total and partial density of states (DOS) with spin-polarization for MnO$_2$ under high-pressure of FM and AFM structures with WIEN2K.

In order to further explore the effect of the rotation octahedrons on the band gaps for *Pnnm* phases, we consider Rutile-phase RuO$_2$, CrO$_2$, SiO$_2$, GeO$_2$, SnO$_2$ and PbO$_2$. They all could be *Pnnm* phase at certain pressure. The band gaps of semi-conductive SiO$_2$, GeO$_2$ and PbO$_2$ are increasing with the rotation of octahedrons, this tendency is in keeping with *Pnnm*-MnO$_2$. RuO$_2$ and CrO$_2$ are conductors or semi-metals[27], though in similar patterns, they are within the scope of discussion. As to SnO$_2$, its bandgap enlarges with increasing pressure, however, we do not find any rotation of the octahedrons. The variation tendency of the bond lengths and bond angles of the octahedrons are shown in Fig. S7.

We find that not only with a good possible PV efficiency, MnO$_2$ could be applied at harsh working conditions for its good mechanical properties. We calculate the bulk modulus (B), shear modulus (G), Young's modulus (Y), B/G ratio and Vickers hardness (Hv) of FM- $P4_2/mnm$, AFM- $P4_2/mnm$, FM- *Pnnm*, AFM- *Pnnm*, FM-$Pa\bar{3}$, AFM-$Pa\bar{3}$ phases strain–stress method, and the results are shown in Table S1. It can clearly see that all the elastic constants C$_{ij}$ satisfy the Born-Huang criterion [44] (Table 2). Thus, the four HP structures are all mechanically stable. Compared with the brittleness of monocrystalline silicon, the structures of *β*- MnO$_2$ under pressure show remarkable ductility. The B/G ratio are all more than the 1.75 critical value [45]. For the inorganic semiconductors, it is extremely difficult to find a ductility favourable material [46]. And it is surprising to see that $Pa\bar{3}$ phase has high bulk modulus of 470 GPa for FM state and 456 GPa for AFM state. Its bulk modulus is even higher than that of a diamond, B = 444 GPa, which is the hardest material known. And the $Pa\bar{3}$-RuO$_2$ with B = 380 GPa [47] has been reported as a potential super hard material. So we study the Vickers hardness for above-mentioned four states with Chen's model [48]. The Vickers hardness is 14 GPa for the AFM-$Pa\bar{3}$ phase. Amongst PV materials, its hardness is much higher than the perovskite solar cells, therefore it could be used in extreme condition.

We calculate the dielectric function which can be expressed as $\varepsilon(\omega) = \varepsilon_1(\omega) + i\varepsilon_2(\omega)$, where the real parts and imaginary parts are represented $\varepsilon_1(\omega)$ and $\varepsilon_2(\omega)$, respectively. From Fig. 5(a), we can find that the static dielectric constants of antiferromagnetic $P4_2/mnm$, *Pnnm* and $Pa\bar{3}$ phases are 11.67, 10.66 and 11.79 at 0 GPa,

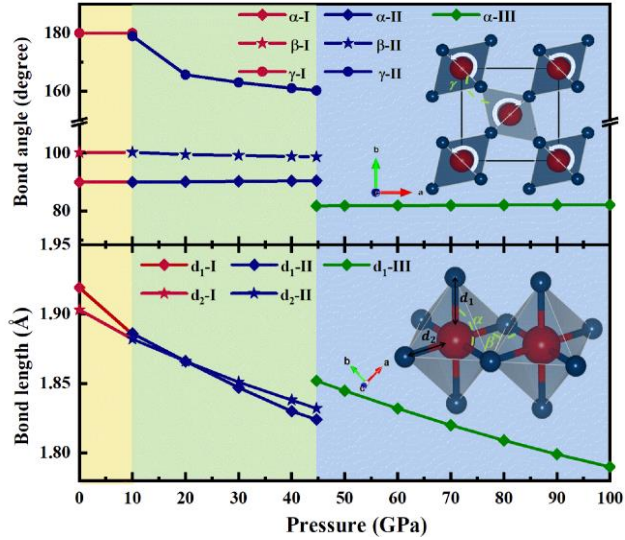

**FIG. 4.** The calculated bond length and bond angle of octahedrons of the $P4_2/mnm$, $Pnnm$, and $Pa\bar{3}$ phases with increasing pressure. The top performs the bond angle of Mn-O-Mn and the inset shows octahedral structure diagram. The bottom shows the bond length of Mn-O and the inset shows the directions of $MnO_6$ octahedrons rotate around the z axis with the pressure increasing and the tetragonal phase distort to orthorhombic phase.

40 GPa and 70 GPa, respectively. The results indicate that the $Pa\bar{3}$ phase has strong polarization and charges binding ability. The imaginary parts link the process of interband transition and the electronic structure. In Fig. 5(b), combined with the analysis of the density of states, we can see that the contribution of the main peaks could be from O-$2p$ orbit to Mn-$d$ and the hybridization orbitals of O-$2p$ and Mn-$3d$ to Mn-$3d$ orbitals of electron transition. It is clearly that the curves of optical properties occur blue shift with the increasing of pressure.

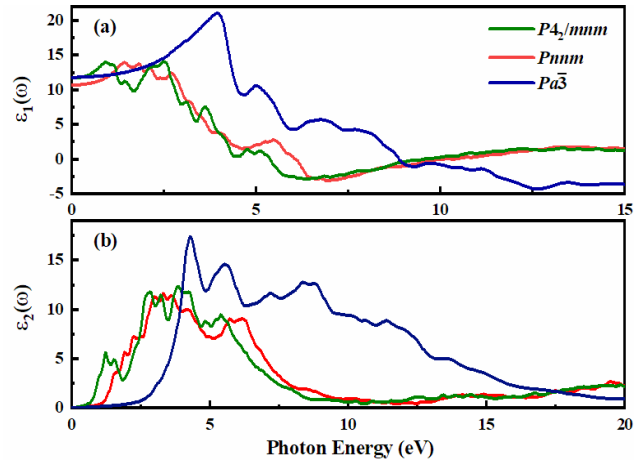

**FIG. 5.** Dielectric function of the $P4_2/mnm$, $Pnnm$ and $Pa\bar{3}$ phases at 0 GPa, 40 GPa and 70 GPa, respectively. (a) Real parts of the dielectric function. (b) Imaginary parts of the dielectric function.

# IV. CONCLUSION

We lauched a sysmatcial study on the octahedron bearing non-Perovskite magnetic $MnO_2$ under 0-100 GPa with DFT. Its phase sequence is decided, and it is always an antiferromagnetic semiconductor. We find that the distortion of octahedra leads to a pressure-induced bandgap enlargement and shrinking in its phase-Ⅱ, due to the competition between bondlength and bonding angle, which gives a clue for modulation and designation of functional material with pressure. Due to the different ways of octahedral connection in phase-Ⅱ and phase-Ⅲ, the bandgap broadens greatly. And the band gap could be tuned to 1.34 eV with pressure in the phase-Ⅲ, which is the best for solar celler's conversion efficiency . Moreover, these two HP phases could be kept to ambient pressure. Mechanically, it is hard and ductile. Together with its optic properties, we conclude that $MnO_2$ could be an excellent candidate for multifunctional PV material in all working condition. We also find similar evolution pattern of bandgaps caused by the distortion of the octahedron in $SiO_2$, $GeO_2$, $SnO_2$ and $PbO_2$. Our study extends $MnO_2$'s application and give new mechanism of pressure induced bandgap evolution. To control either the bond lengths or the bond angles meight be a effective way of the electronic structure designation.

# ACKNOWLEDGEMENS

This work was supported by the National Key R&D Program of China (No. 2018YFA0703404, 2016YFB0201204, 2017YFA0403704), National Natural Science Foundation of China (Nos. 91745203, 11774121). Program for Changjiang Scholars and Innovative Research Team in University (No. IRT_15R23), Parts of calculations were performed in the High Performance Computing Center (HPCC) of Jilin University.

# REFERENCES

[1] Y. Zhang, J. Wang, and P. Ghosez, Physical Review Letters **125** (2020).
[2] Y. Wang, F. Tian, D. Li, D. Duan, H. Xie, B. Liu, Q. Zhou, and T. Cui, J. Alloys Compd. **788**, 905 (2019).
[3] Y. Liu, D. Li, F. Tian, D. Duan, B. Liu, and T. Cui, Inorganic Chemistry Frontiers **7**, 1108 (2020).
[4] D. Liu and T. L. Kelly, Nat. Photonics **8**, 133 (2013).
[5] B. D. Alexander, P. J. Kulesza, I. Rutkowska, R. Solarska, and J. Augustynski, J. Mater. Chem. **18**, 2298 (2008).
[6] L. Zhang, Y. Wang, J. Lv, and Y. Ma, Nature Reviews Materials **2** (2017).
[7] S. Carenco, D. Portehault, C. Boissiere, N. Mezailles, and C. Sanchez, Chem. Rev. **113**, 7981 (2013).
[8] K. Zhao, S. Gupta, C. Chang, J. Wei, and N.-H. Tai, RSC Adv. **9**, 19217 (2019).
[9] Y. Duan, H. Pang, Y. Zhang, J. Chen, and T. Wang, Mater. Charact. **112**, 206 (2016).
[10] Y. Duan, Y. Zhang, J. Chen, Z. Liu, and T. Wang, Mater. Chem. Phys. **157**, 1 (2015).
[11] Z. Zhang, S. Wang, Y. Lv, X. Chen, Z. Wu, and Y. Zou, J. Alloys Compd. **810**, 151744 (2019).
[12] W. L. Song, M. S. Cao, Z. L. Hou, M. M. Lu, C. Y. Wang, J. Yuan, and L. Z. Fan, Appl. Phys. A: Mater. Sci. Process. **116**, 1779 (2014).
[13] D. Wang, L. M. Liu, S. J. Zhao, B. H. Li, H. Liu, and X. F. Lang, Phys. Chem. Chem. Phys. **15**, 9075 (2013).
[14] F. H. B. Lima, M. L. Calegaro, and E. A. Ticianelli, Electrochim. Acta **52**, 3732 (2007).
[15] J. Zang and X. Li, J. Mater. Chem. **21**, 10965 (2011).
[16] A. Chartier, P. D'Arco, R. Dovesi, and V. R. Saunders, Phys. Rev. B **60**, 14042 (1999).
[17] X. Hu, J. Chen, S. Li, Y. Chen, W. Qu, Z. Ma, and X. Tang, The Journal of Physical Chemistry C **124**, 701 (2019).
[18] X. Huang, X. H. Yan, Y. Xiao, Y. D. Guo, Z. H. Zhu, and C. J. Dai, Europhys. Lett. **99**, 27005 (2012).
[19] M. Kan, J. Zhou, Q. Sun, Y. Kawazoe, and P. Jena, J. Phys. Chem. Lett. **4**, 3382 (2013).
[20] H. Sato, T. Enoki, M. Isobe, and Y. Ueda, Phys. Rev. B **61** (2000).
[21] J. P. S. R. DRUILHE, Czech. J. Phys. B **17**, 337 (1967).
[22] S. L. Dudarev, G. A. Botton, S. Y. Savrasov, C. J. Humphreys, and A. P. Sutton, Phys. Rev. B **57**, 1505 (1998).
[23] J. Haines, J. M. Léger, and S. Hoyau, J. Phys. Chem. Solids **56**, 965 (1995).
[24] L. G. Liu, Earth Planet. Sci. Lett. **29**, 104 (1976).
[25] L. S. Dubrovinsky, N. A. Dubrovinskaia, V. Swamy, J. Muscat, N. M. Harrison, R. Ahuja, B. Holm, and B. Johansson, Nature **410**, 653 (2001).
[26] Y. Li, W. Fan, H. Sun, X. Cheng, P. Li, X. Zhao, J. Hao, and M. Jiang, J. Phys. Chem. A **114**, 1052 (2010).
[27] V. Srivastava, M. Rajagopalan, and S. P. Sanyal, The European Physical Journal B **61**, 131 (2008).
[28] J. Haines, J. M. Léger, C. Chateau, R. Bini, and L. Ulivi, Phys. Rev. B **58**, R2909 (1998).
[29] A. Togo, F. Oba, and I. Tanaka, Phys. Rev. B **78** (2008).
[30] A. R. Oganov, A. O. Lyakhov, and M. Valle, Acc. Chem. Res. **44**, 227 (2011).
[31] G. Kresse and J. Furthmüller, Physical Review B Condensed Matter **54**, 11169 (1996).
[32] H. J. Monkhorst and J. D. Pack, Phys. Rev. B **13**, 5188 (1976).

[33] R. Hill, Proceedings of the Physical Society **65**, 349 (1952).

[34] A. I. Liechtenstein, V. I. Anisimov, and J. Zaanen, Phys. Rev. B **52**, R5467 (1995).

[35] C. Franchini, R. Podloucky, J. Paier, M. Marsman, and G. Kresse, Phys. Rev. B **75** (2007).

[36] A. A. Bolzan, C. Fong, B. J. Kennedy, and C. J. Howard, Aust. J. Chem. **46**, 939 (1993).

[37] M. Regulski, R. Przeniosło, I. Sosnowska, and J. U. Hoffmann, Phys. Rev. B **68** (2003).

[38] J. S. Lim, D. Saldana-Greco, and A. M. Rappe, Phys. Rev. B **94** (2016).

[39] K. Schwarz and P. Blaha, Comput. Mater. Sci. **28**, 259 (2003).

[40] E. R. Ylvisaker and W. E. Pickett, Phys. Rev. B **79**, 5103 (2009).

[41] D. A. Kitchaev, H. Peng, Y. Liu, J. Sun, J. P. Perdew, and G. Ceder, Phys. Rev. B **93** (2016).

[42] Y. Liang *et al.*, Adv. Sci., 1900399 (2019).

[43] T. Yin *et al.*, J. Am. Chem. Soc. **141**, 1235 (2019).

[44] Z. J. Wu, E. J. Zhao, H. P. Xiang, X. F. Hao, X. J. Liu, and J. Meng, Phys. Rev. B **76** (2007).

[45] S. F. Pugh, The London, Edinburgh, and Dublin Philosophical Magazine and Journal of Science **45**, 823 (2009).

[46] T.-R. Wei *et al.*, Science **369**, 542 (2020).

[47] J. S. Tse, D. D. Klug, K. Uehara, Z. Q. Li, J. Haines, and J. M. Léger, Phys. Rev. B **61**, 10029 (2000).

[48] X. Q. Chen, H. Niu, D. Li, and Y. Li, Intermetallics **19**, 1275 (2011).